\documentclass[twocolumn,showpacs,preprintnumbers,amsmath,amssymb, nofootinbib]{revtex4}

\usepackage{pdflscape}
\usepackage{color}
\usepackage{dcolumn}
\usepackage{bm}
\usepackage{epsfig}%
\usepackage{amsmath}
\usepackage[colorlinks]{hyperref}
\usepackage{ulem}

\newcommand{\tr}{\textcolor{black}}

\begin{document}

\title{Electrodynamics of superconductors}

\author{V. Kozhevnikov }
\affiliation{Tulsa Community College, Tulsa, Oklahoma 74119, USA}.  \\
\affiliation{KU Leuven, BE-3001 Leuven, Belgium}.  \\


\begin{abstract}
\noindent Electrodynamics of superconductors is primarily the electrodynamics of the Meissner state, a state characterized by zero magnetic induction of a superconducting fraction of conduction electrons. Simultaneously, the Meissner state is characterized by zero resistivity and zero entropy of these electrons. The latter means that the temperature of an ensemble of superconducting electrons is zero as well. Understanding properties of the Meissner state provides a key to understand and predict these and other equilibrium and non-equilibrium properties of superconductors, e.g., properties of the intermediate and mixed states, flux quantization, resistanceless transport current, etc. The reader will see that all these properties have a single and very simple origin: quantization of the angular momentum of the conduction electrons combined into Cooper pairs, and that the electrodynamics of superconductors and superconductivity as a whole is a magnificent manifestation of the boundless ingenuity of Nature and of the power of Physics laws.  
 
\end{abstract}\

\maketitle  
\vspace{3mm}

\textbf{Keypoints}

$\bullet$ The electromagnetic and thermal properties of superconductors are dictated by  Cooper pairs. On this reason, the electrodynamics of superconductors is primarily the electrodynamics of an ensemble of Cooper pairs.

$\bullet$ By definition of Cooper pairs, the paired electrons orbit their center of mass. The stability of Cooper pairs means that the pairs obey the Bohr-Sommerfeld quantization condition. 

$\bullet$ In the Meissner phase (the phase of a superconducting sample with zero magnetic induction $B$) the magnetic field intensity $H$ is equal to or greater than the intensity of the applied field $H_0$.   

$\bullet$ The properties of superconductors (zero resistivity, zero magnetic induction, zero entropy, flux quantization, \textit{etc.}) are governed by the precession of Cooper pairs with  quantized angular momentum. 

$\bullet$ Due to zero entropy, the temperature of an ensemble of Cooper pairs equal to zero regardless of the temperature of the sample.  

\section*{INTRODUCTION}

Superconductivity was discovered in the laboratory of Kameringh Onnes in 1911 as a phenomenon of a sharp drop of electrical resistance in a high purity mercury wire at temperature $T$ below 4.2 K \cite{Onnes_1911}.  An original estimate of the resistance drop was 10$^4$ times compare to the resistance just above the transition temperature, called the critical temperature $T_c$, a constant of the superconducting (S) material. Fairly soon it became clear that resistance of a superconductor is not merely a small, but zero. Namely, the current in a closed S circuit represents a persistent current  like that caused by electrons bound in atoms\footnote{Persistent current is a current running in a closed circuit without both  thermal and radiation losses. }. It was also found that the S state exists up to a certain temperature-dependent critical magnetic field $H_c(T)$,  which is another characteristic of the material. While number of superconductors was (and is) steadily growing, at an early stage it seemed  that zero resistivity is the only property distinguishing superconductors from the so-called normal (N) metals. 

The second hallmark of superconductivity, the disappearance of thermoelectric effects, was discovered by Walter Meissner in 1927 \cite{Meissner27}. Lack of thermo-e.m.f.  suggests that entropy of the S fraction of conduction electrons (called ``superconducting electrons") is zero. If so, then superconductivity can be a previously unknown thermodynamic state of matter characterized by complete ordering of superconducting electrons. Gorter and Casimir used zero entropy as the base postulate of their two-fluid model (1934) \cite{Gorter_Casimir}, a seminal thermodynamic theory addressing properties of superconductors in zero field. On the other hand, a non-zero thermo-e.m.f. looks incompatible with zero resistivity of the S state, since otherwise it would lead to an infinite current and therefore to an instantaneous restoration of the N state.   

The idea of the new thermodynamic state was brilliantly confirmed in a so-called Meissner effect independently discovered by Meissner and Ochsenfeld\footnote{An active promoter of this experiment was von Laue.} (1933) \cite{Meissner} and Ryabinin and Shubnikov (1934) \cite{Shubnikov}. It was found that, apart from zero resistivity and zero entropy, superconductivity is also characterized by zero magnetic induction $B$ regardless on the history of the field application. Hence, a sample in the  Meissner state (MS) represents a perfect diamagnetic, i.e., its magnetic susceptibility $\chi$ and permeability $\mu_m(=1+4\pi\chi)$ are equal to $-1/4\pi$ and zero, respectively. Important to note that (i) like in regular diamagnetics, in the MS $\chi$ and therefore $\mu_m$ do not depend on the sample temperature $T$ and the applied field $H_0$; and (ii) if the first two ``big zeroes" of superconductivity take place in samples of any shape, connectivity and purity, the third zero ($B=0$ or the Meissner effect) is observed only with sufficiently pure singly connected samples of an ellipsoidal shape. 

It was found that there are two kinds of superconductors distinguishing  by the sign of the S/N interphase energy also called surface tension. Materials with positive and negative surface tension are referred to as type-I and type-II superconductors, respectively. All type-I superconductors are elementary metals, while type-II materials, with very few exceptions, are alloys and multi-component compounds. A sample of type-I material can be made behaving like type-II superconductor by introducing impurities and/or reducing the sample size, e.g., by decreasing the thickness of a film sample. However, the opposite is not possible. Other really astonishing properties, such as the flux quantization, Josephson effect and others, were discovered in 1960s and later. 

A search for the source of the superconductivity phenomenon culminated in a ground braking theoretical discovery of Leon Cooper (1956) \cite{Cooper},  who shown that even a weak attraction can lead to formation of stable electron pairs called Cooper pairs. The origin of attraction lays in the medium polarization occurring due to electron-lattice interaction, as it was first suggested by Fr\"{o}hlich (1950) \cite{Frohlich}
and discussed in the BCS theory (Bardeen, Cooper and Schrieffer, 1957) \cite{BCS}. Cooper's prediction was irrefutably confirmed in experiments (Deaver and Fairbank, 1961 \cite{Deaver}; and many others). 

Cooper pairs profoundly alter the medium electrodynamics. The physics behind these alterations is the subject of this chapter. 

The chapter consists of three sections: Meissner state, Flux quantization and Total current, and a brief summary. References to literature for further reading are given at the end. The \textit{cgs}  units are used because these 
units are the most appropriate for discussing  electrodynamics of continuous media\footnote{Different dimensions of the induction $\textbf{B}$ and the intensity $\textbf{H}$ of the magnetic field in the SI unit system, make this system in its current form inapplicable for this purpose.}. In view of the space limit only key references are provided.

\section*{MEISSNER STATE}

\textbf{Definitions.} The MS is an equilibrium S state defined as the state with $B=0$ throughout the sample volume $V$. Alike, the MS can be defined as the state with $\mu_m=0$ or $\chi=-1/4\pi$ all over $V$. 

The MS takes place in sufficiently pure massive and simply connected samples of an ellipsoidal shape in the applied field $H_0<H_{c1}(1-\eta)$. Here $H_{c1}$ is the lower critical field of type-II superconductors equal to the thermodynamic critical field $H_c$ in the type-I counterparts, and $\eta$ is the demagnetizing factor\footnote{The demagnetizing factors are determined by relationship of the ellipsoidal axes. Respectively, $\eta$-s are defined only for the ellipsoidal bodies.} with respect to an ellipsoidal axis parallel to $\textbf{H}_0$. A sample is considered as massive if its minimal size significantly exceeds a so-called penetration depth $\lambda$, a width of the surface layer where $B$ decays from the induction of external field near the sample surface $B_{ext}$ to zero in the sample interior. 

A unique feature of the ellipsoidal bodies in a static magnetic field  is uniformity of the field intensity $\textbf{H}$ (also called the magnetizing force, field strength, etc.)  throughout their volume, as it was first shown by Poisson \cite{Poisson} and discussed by  Maxwell \cite{Maxwell}. The intensity $\textbf{H}$ is responsible for magnetization $\textbf{I}(=\chi \textbf{H})$. Depending on a demagnetizing field $\textbf{H}_d(=4\pi\eta \textbf{I}$), in diamagnetics  $\textbf{H}(=\textbf{H}_0-\textbf{H}_d)$ is greater than or equal to $\textbf{H}_0$ in magnitude, and its direction can differ from that of the applied field. $\textbf{H}$ is the field acting on ``native" (belonging to the body) charges, whereas the induction $\textbf{B}$ is the field acting on a ``foreign" (extraneous) charges\footnote{On that reason the field measured by a prob charge, e.g., in $\mu$SR, is $B$, whereas in magnetic resonances the measured or controlling field is $H$. }. Contrarily to $\textbf{B}(=\textbf{H}+4\pi \textbf{I})$, the average of  microscopic fields caused by all native charges, $\textbf{H}$ does not include the field due to the charge in question\footnote{Here, the charge should be understood as the current caused by all types of microscopic movement, including the spinning one, of a given charge. }. Respectively, in any medium $\textbf{H}$ differs from $\textbf{B}$. On the contrary, in a free space (vacuum) $\textbf{I}=0$ and, consequently, $\textbf{\textbf{H}}=\textbf{\textbf{B}}$\footnote{Important that this is not just a mathematical coincidence, but a physical identity. This implies that in any system of physical units the dimensions of $\textbf{B}$ and $\textbf{H}$ must be the same.}. 

At the sample boundary the normal component of the intensity $H_n$ undergoes a discontinuity $\Delta H_n=4\pi I_n$, where $I_n$ is the magnitude of the normal component of magnetization $\textbf{I}$. Correspondingly, the intensity can be viewed as a potential field created by magnetic charges sitting on the boundary and having a surface density $I_n$. On the other hand, the intensity inside the sample, i.e. in any volume element $dV$ not crossed by the sample boundary, represents a solenoidal or divergenceless field, which, therefore, can be described by a vector potential.  
As an example, Fig.\,1 schematically shows the induction $\textbf{B}$ and the intensity $\textbf{H}$ of the magnetic field in and out of a sample with $\eta=1/2$ when it is in the MS\footnote{The London theory, adopted for description of the MS in the Ginzburg-Landau and BCS theories, is based on an assumption that in superconductors the magnetic susceptibility $\mu_m$ and the dielectric permittivity $\varepsilon_e$ are unity and therefore $\textbf{B}=\textbf{H}$. However, this is only true in a vacuum.  }.
\begin{figure}
	\includegraphics[width=0.85\linewidth]{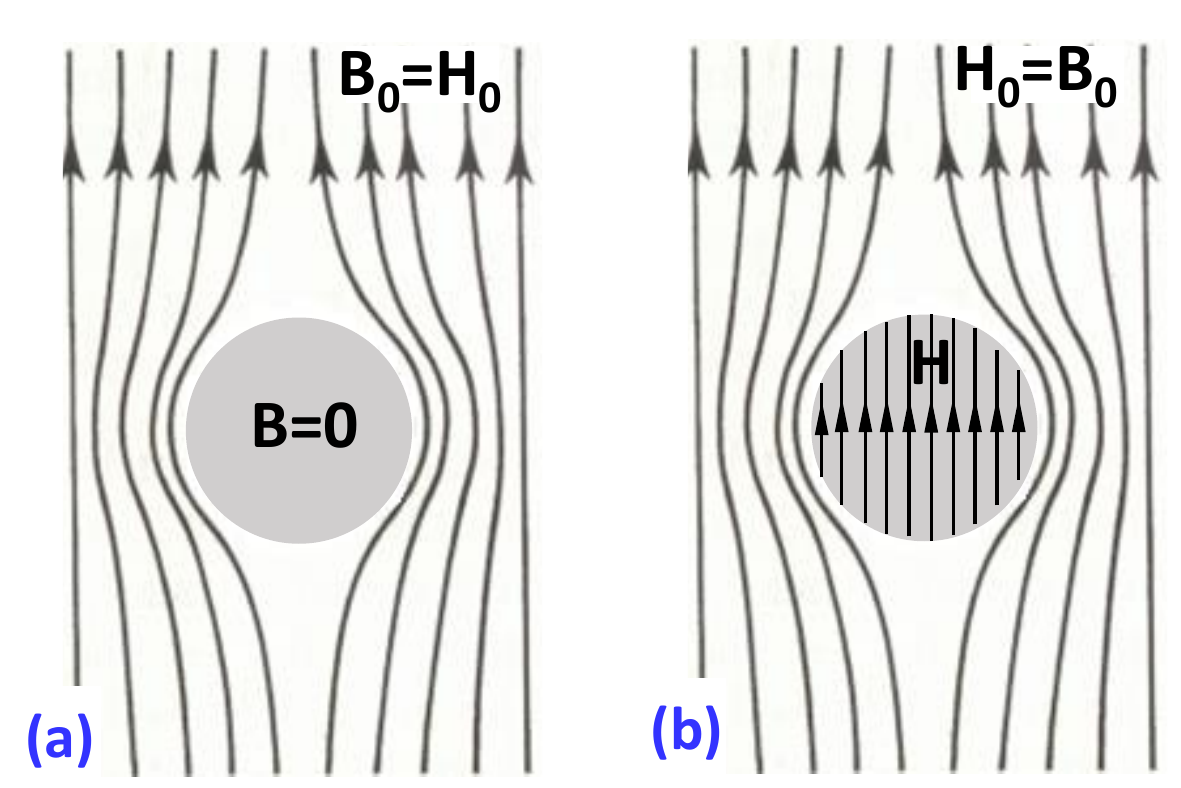}
	\caption{Induction $\textbf{B}$ (a) and intensity $\textbf{H}$ (b) of the magnetic field inside and outside of a long cylindrical sample in the Meissner state when the field $\textbf{H}_0$ is applied transversely to its longitudinal axis. For this sample/field configuration $\eta=1/2$. Inside the sample  $\textbf{B}=0$ and $\textbf{H}=\textbf{H}_0/(1-\eta)=2\textbf{H}_0$; outside it $\textbf{H}=\textbf{B}$ and far away $\textbf{H}_0=\textbf{B}_0$. Shadowed circle is a cross-section of the sample oriented perpendicular to the page.  }
	\label{fig:epsart}
\end{figure}

If $\textbf{H}_0$ is parallel to an axis of the ellipsoidal sample in the MS, $\textbf{H}=\textbf{H}_0/(1-\eta)$, where $\eta$ is the demagnetizing factor about this axis. If $\textbf{H}_0$ is not parallel to either axis, $\textbf{H}$ is the vector sum of the intensities  relative to each axis. At any orientation of $\textbf{H}_0$ with respect to the sample in the MS, inside it the magnitude of intensity $\textbf{H}$ is less than $H_{c1}$.

After all, a very important property of the MS follows from the Law of the symmetry of time reversal, mandatory for any (either classical or quantum) equilibrium state. According to this Law, no total current can occur in the equilibrium state\footnote{The time transformation $t\rightarrow -t$ changes the direction of  magnetic moment of a system with total current and therefore such a system can not be in the equilibrium state by definition. } and therefore all currents in the MS must be compensated. 

In samples with $\eta=0$, referred to as the samples of cylindrical geometry, the outer field due to the magnetized sample is absent and respectively $\textbf{H}=\textbf{H}_0$ both inside and outside the sample. Such samples can be, e.g., long cylinders, infinite slabs or wide ribbon-like foils in $\textbf{H}_0$ parallel to their generating  line\footnote{Any sample with $\eta=0$ can be broken for a large number of identical thin circular cylinders parallel to $\textbf{H}_0$.}.  
Here our discussion of the MS will be limited  by the cylindrical samples. 
\begin{figure}
	\includegraphics[width=0.9\linewidth]{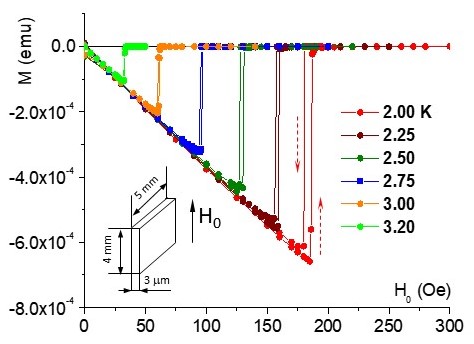}
	\caption{Magnetic moment of a 2.9-$\mu$m thick indium film in the parallel field $\textbf{H}_0$ measured at constant temperatures, as indicated. For this sample $\eta=0$ and $\textbf{H}=\textbf{H}_0$. The data were taken with the sample cooled in zero field (arrow up for 2.0 K) and in the  field exceeding $H_c$ (arrow down for 2.0 K). Hysteresis at the S/N transition is the supercooling effect caused by the positive S/N surface tension. The film geometry is shown in the insert. (After V. Kozhevnikov et al. JSNM \textbf{33}, 3361 (2020).)} 
	\label{fig:epsart}
\end{figure}

In this case the magnetic moment $\textbf{M}$ of the sample is
\begin{equation}
	\textbf{M}\equiv\chi V\textbf{H}=-\frac{V}{4\pi}\textbf{H}_0
\end{equation}

and the sample magnetic energy $E_m$ is
\begin{equation}
	E_m\equiv-\int \textbf{M}\cdot d\textbf{H}_0=-\frac{\textbf{M}\cdot\textbf{H}_0}{2}=\dfrac{VH_0^2}{8\pi}.
\end{equation}

By virtue of the energy conservation, in the cylindrical samples $E_m=\Delta T$, where the latter is the field-induced change of the kinetic energy of electrons. 
As an example, graphs for $\textbf{M}$ vs $\textbf{H}_0$ for a type-I superconductor  with $\eta=0$ are shown in Fig.\,2.

\textbf{Cooper pairs}. The Cooper pairs are defined as a correlated state of two conduction electrons with zero net kinetic linear momentum with respect to their center of mass and zero net spin. The latter condition stems from the requirement of thermodynamics since the free energy of a system of units with zero spin  is lesser than it would be if the spin is not zero. The same condition justifies the thermodynamic advantage of the electron pairing since it allows to null the spin.

Zero net kinetic linear momentum and stability of the pairs mean that the paired electrons orbit their center of mass like, e.g., stars in a  binary star. Therefore, each Cooper pair possesses the angular momentum $\bm{\iota}$ and the magnetic moment $\bm\mu=\gamma\bm{\iota}$ , where $\gamma$ is the gyromagnetic ratio. As measured by I. Kikoin and Goobar \cite{Isaak-1}, $\gamma$ in superconductors equals its classical value $e/2mc$. Here $m$ and $e$ are mass and charge of one electron, respectively, and $c$ is the electromagnetic constant of the \textit{cgs} unit system equal to speed of light. This confirms that the ``superconducting" electrons are effectively spinless.  

Stability of the pairs also means that the orbital motion of the paired electrons occurs without energy dissipation. Therefore they must obey the Bohr-Sommerfeld quantization condition. At the same time, one should bear in mind that  to meet the law of momentum conservation,  in a magnetic field the linear momentum should be taken in a generalized form. Hence, the quantization condition for the single Cooper pair is
\begin{equation}
\oint \widetilde{\textbf{p}}_{cp}\cdot d\textbf{l}=2\pi R\widetilde{p}_{cp}=2\pi \widetilde{\iota}_{cp}= nh,
\end{equation}
where $\widetilde{p}_{cp}$ and $\widetilde{\iota}_{cp}$ are the generalized linear and angular momentum of the pair, respectively, $R$ is the radius of orbital motion of paired electrons, $h$ is the Planck constant, and $n$ is a non-negative integer ($n=0, 1, 2, ...$). 

By definition of $\widetilde{\textbf{p}}$, the generalized linear momentum of the paired electrons  $\widetilde{\textbf{p}}_{cp}$ is 
\begin{equation}
	\widetilde{\textbf{p}}_{cp}=\widetilde{\textbf{p}}_1+\widetilde{\textbf{p}}_2=(m\textbf{v}_1+\frac{e\textbf{A}_1}{c})+(m\textbf{v}_2+\frac{e\textbf{A}_2}{c}),
\end{equation}
where $\textbf{v}$ is the electron velocity and $\textbf{A}$ is the vector potential of the field acting on the electron, i.e. of the intensity $\textbf{H}$; subscripts $1$ and $2$ denote the quantities related to the first and second electron in the pair. 
\begin{figure}
	\includegraphics[width=0.7\linewidth]{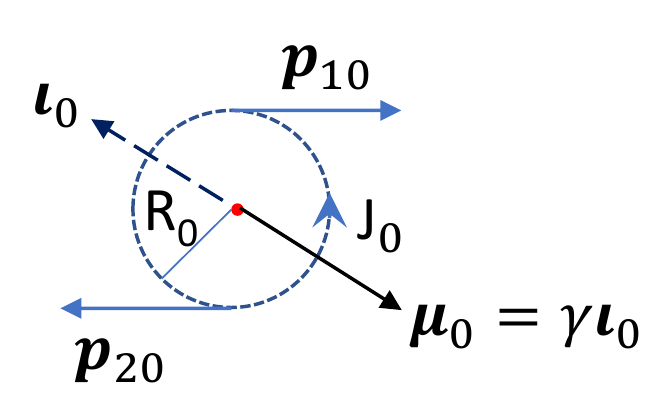}
	\caption{Schematics of a single Cooper pair at zero field and temperature. $\bm{\iota}_0$ and $\bm{\mu}_0$ are the angular momentum and the magnetic moment of the pair, respectively; $\gamma$ is the gyromagnetic ratio;  $\textbf{p}_{10}$ and $\textbf{p}_{20}$ are the kinetic linear momentums of electrons; $J_0$ is the current caused by the orbital motion of electrons round their center of mass; $R_0$ is the orbit radius; red dot designates the center of mass of the pair. } 
	\label{fig:epsart}
\end{figure}

In the ground state, i.e. at zero field and temperature, $n$ in Eq.\,(3) takes the lowest value $n=0$ and $\widetilde{\textbf{p}}_{cp}$ is the sum of  kinetic linear momentums $\textbf{p}(=m\textbf{v})$ of electrons. Hence,
\begin{equation}
	(\widetilde{\textbf{p}}_{cp})_0=(\textbf{p}_{cp})_0=\textbf{p}_{10}+\textbf{p}_{20}=0,
\end{equation}
where subscript $0$ denotes the zero field. 

The last part of Eq.\,(5) is identical to the definition of Cooper pair.  The first part of this equation [($\widetilde{\textbf{p}}_{cp})_0=0$] is a modified London's rigidity, originally formulated for a single superconducting electron \cite{London50}. Also, Eq.\,5 implies that the center of mass of the pair is at rest with respect to the sample.

The motion of electrons in one Cooper pair at zero field and temperature is schematically shown in Fig.\,3. The orbit diameter $2R_0$ corresponds to the correlation length $\xi_0$ of the BCS theory and $\xi$ of the Ginzburg-Landau (GL) theory \cite{GL}, which are close to each other at these conditions. 

One more important property at zero field is related to the ensemble of Cooper pairs: due to symmetry, the total magnetic moment of all pairs (the magnetic moment of the sample) is zero, i.e. 
\begin{equation}
	\textbf{M}_0=\sum\bm{\mu}_0=0,
\end{equation}
where summation is taken over all pairs\footnote{Note that Eq.\,(6) resembles the definition of a diamagnetic atom, where each electron has a non-zero orbital momentum, while the momentum of the atom is zero. }.

In view of uniformity of the bulk properties of the MS, Eq.\,(6) holds for the unit volume as well as for a physically infinitesimal volume element $dV$. 
The latter in superconductors is defined as a volume, which size is much smaller than the size of the volume taken by the S phase and much larger than the spacial inhomogeneity of microscopic currents, i.e. $R_0$. We remind that Cooper pairs are heavily overlap, but it affects neither stability, no mobility of the pairs. Taking into account the wave side of the electron nature, this fact is similar as the overlapping of myriads of electromagnetic waves does not prevent us from seeing objects or enjoying music broadcast from the other side of the globe. 

Now it is rather obvious what happens when the magnetic field is turned on: Cooper  pairs precess. Let us see how does it go. 

In the field, each of the paired electrons (as well as all other charges constituting the sample) experiences the action of the Lorentz force\footnote{The action of a magnetic field on atomic electrons and nuclei results in the usual diamagnetic response; its action on unpaired conduction electrons causes a weak Landau diamagnetism.}. In absence of the applied electric field\footnote{A static electric field does not penetrate the superconducting sample due to the screening effect of always present ``normal" electrons.},  this force is  
\begin{multline}
	\textbf{F}\equiv\frac{d \textbf{p}}{d t}=-\frac{e}{c}\left(\frac{\partial \textbf{A}}{\partial t}\right)+\frac{e}{c}\textbf{v}\times \textbf{H}=\\-\frac{e}{c}\left(\frac{\partial \textbf{A}}{\partial t}\right)+\frac{e}{c}\textbf{v}\times (\nabla \times \textbf{A}). 
\end{multline}

Here $t$ is time, $\textbf{A}$ is the vector potential  of the field $\textbf{H}(=\nabla \times \textbf{A})$, and $\textbf{v}$ is the electron velocity, which magnitude is about the same as the Fermi velocity $v_F$ (the maximum difference between $v$ and $v_F$ is on the order of a percent).  

Important to note that since inside real bodies $\textbf{B}$ and $\textbf{H}$ are different, one has to distinguish the vector potential for $\textbf{H}$ and for $\textbf{B}$. In our notations $\textbf{\textbf{A}}$ is the vector potential of the intensity $\textbf{H}$ inside the sample. Below we will see how $\textbf{A}$ is related to the vector potential $\textbf{A}_B$ defined by the relationship $\textbf{B}=\nabla\times \textbf{A}_B$. As usual, the  supplementary condition $\nabla\cdot \textbf{A}=0$ is presumed.

The first term in  the right-hand side of Eq.\,(7) is an electric force $\textbf{F}_E$ caused by the changing  magnetic field.  Corresponding vortex electric field $\textbf{E}$ is defined as
\begin{equation}
	\textbf{E}\equiv \frac{\textbf{F}_E}{e} =-\frac{1}{c}\left(\frac{\partial \textbf{A}}{\partial t}\right).
\end{equation}

As with bound electrons in conventional materials, the electric field $\textbf{E}$ (existing while the magnetic field is changing) does the work resulting in a change of kinetic energy of electrons and  in the appearance  of an induced magnetic moment in each pair.

Specifically, when the vector potential changes from zero to $\textbf{A}$, which corresponds to the intensity change from zero  
to $\textbf{H}$, the velocity of electrons in the pair is changed from $\textbf{v}_0$ to $\textbf{v}'$. The difference $\textbf{v}_i= \textbf{v}'-\textbf{v}_0$ is the velocity induced due to the applied magnetic field. Integrating Eq.\,(8) over time, one obtains 
\begin{equation}
	\textbf{v}_i=-e\frac{\textbf{A}}{cm}. 
\end{equation}
\begin{figure}
	\includegraphics[width=0.7\linewidth]{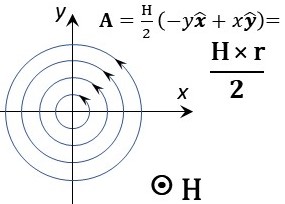}
	\caption{Lines of the vector potentials  $\textbf{A}$ of the circular gauge for a uniform magnetic field $\textbf{H}$ directed toward the reader (along the $z$-axis); $\hat{\bm{x}}$ and $\hat{\bm{y}}$ are unit vectors along $x$ and $y$ axes, respectively. $\textbf{r}$ is a radius vector lying in $xy$ plane. As for all vector fields,  the line of $\textbf{A}$ is the directional line tangential to  the vector $\textbf{A}$ in each of its point.} 
	\label{fig:epsart}
\end{figure}

This is a very important formula. It shows that $\textbf{v}_i$ is parallel to $\textbf{A}$. Therefore, since by definition $\textbf{A}$ lies in the plane perpendicular to $\textbf{H}$, $\textbf{v}_i$ and therefore the induced current also lies in the plane transverse to $\textbf{H}$. On the other hand, the induced current makes a closed loop (it cannot leave the sample) and the loop must be circular due to rotational symmetry stemming from uniformity of $\textbf{H}$ (all directions in the transverse plane are equivalent); therefore lines of the  vector potential make the circular loops either. Hence, an appropriate gauge for the vector potential of $\textbf{H}$ is the circular one, namely $\textbf{A}=\textbf{H}\times \textbf{r}/2$. The lines of this vector potential are shown in Fig.\,4. 

Thus, since $e<0$, Eq.\,(9) indicates that the current induced in each Cooper pair creates a magnetic moment directed opposite to $\textbf{H}$, i.e. the induced moment is always diamagnetic, in accord with requirements of thermodynamics and the time reversal symmetry. This also means that diamagnetism (both in normal and superconducting materials) results from the changing magnetic field in full consistency with the laws of the classical physics. 

Next, like in regular diamagnetics, in Eq.\,(9) the time has dropped out implying that the induced velocity $\textbf{v}_i$ is indifferent to the rate of the field change. 

After all, Eq.\,(9) implies that the vector potential in superconductors is not gauge-invariant since a change in $\textbf{A}$ changes the induced moment and because $\textbf{A}$ of any but the circular gauge does not possess the necessary rotational symmetry. However, this is not  surprising\footnote{Recall that the gauge-invariance does not hold in the London and BCS theories.},  since the gauge invariance is inapplicable to quantities defined by the vector potential in explicit form.  Non-fulfillment of the gauge-invariance in superconductors  is an additional confirmation of the fact that the vector potential is a real and primary characteristics of the magnetic field, as demonstrated by the  Aharonov-Bohm effect \cite{Aharonov_Bohm_1959}. 

\begin{figure}
	\includegraphics[width=0.8\linewidth]{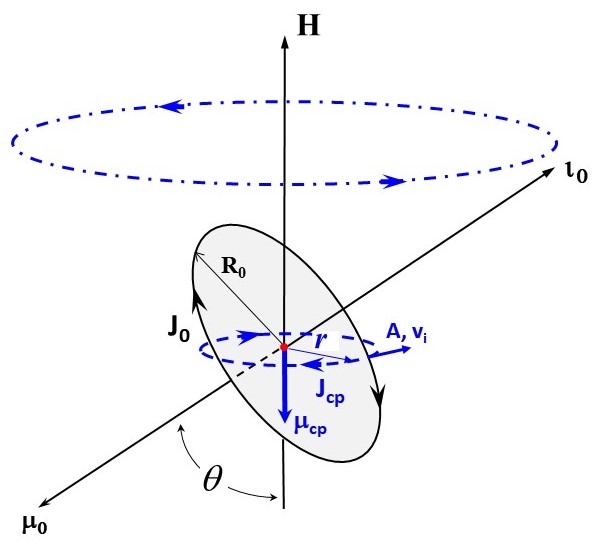}
	\caption{Schematics of the Cooper pair precession in the magnetic field $\textbf{H}$. Vectors $\bm{\iota}_0$ and $\bm{\mu}_0$, current $J_0$ and radius $R_0$ are the same quantities as those at zero field shown in Fig.\,3. A dot-dashed circle designates the path of the tip of the precessing $\bm{\iota}_0$. A dashed blue circle depicts the  field-induced current $J_{cp}$; it is also a line of the vector potential directed opposite to the current $J_{cp}$. $\textbf{A}$ and $\textbf{v}_i$ designate the vector potential and the induced velocity of one electron, respectively; $r$ is the radius of the induced current; $\bm{\mu}_{cp}$ is the induced magnetic moment of the pair. Designated by a single arrow $\textbf{A}$ and $\textbf{v}_i$ are proportional but not equal. After V. Kozhevnikov, JSNM \textbf{34}, 1979 (2021).} 
	\label{fig:epsart}
\end{figure}

The second term in the right hand side of Eq.\,(7) is termed the magnetic  force $\textbf{F}_M$. It is perpendicular to the electron velocity and therefore does no work. Hence, $\textbf{F}_M$ does not affect the electron kinetic energy and the magnitude of its magnetic moment. On the other hand, since the uniform magnetic field cannot change position of the pair's center of mass, $\textbf{F}_{M1}=-\textbf{F}_{M2}$ (those are the magnetic forces acting on the 1st and 2nd electron in the pair),  and  also $\textbf{F}_{M1}$ and $\textbf{F}_{M2}$ are non-central forces. Hence, they make a couple resulting, due to non-zero $\bm{\iota}_0$, in precession of $\bm{\mu}_0$ relative to the vector $\textbf{H}$, as schematically shown in Fig.\,5.

The angular velocity of precession, referred to as Larmor frequency, is 
\begin{equation}
	\bm{o}=-\gamma \textbf{H}=-\frac{eg_L}{2mc}\textbf{H},
\end{equation}
where $g_L$ is the Lande factor. As mentioned, in superconductors $\gamma$ takes the classical value, i.e. $g_L=1$.  

As follows from the Larmor theorem, if $v_i\ll v_0$, precession of the electron orbit is equivalent to the undisturbed orbital motion at the field absence (i.e. with fixed $\bm{\mu}_0$  and $R_0$) plus an additional (field induced) circular motion with the angular velocity $\bm{o}$ and radius $r$ proportional to $R_0$, which leads to appearance of the diamagnetic moment $\bm{\mu}_{cp}$. On the other hand, the invariability of $\bm{\mu}_0$ means that Eq.\,(6) is valid both in the absence and presence of the magnetic field. This can be also understood basing on the fact that precession is an inertia-free motion and therefore all pairs precess synchronously\footnote{Also, as it is shown below, the induced circular motion of the paired electrons occurs in phase.}, i.e. without changing their mutual orientation. 

Thus, we arrive to conclusion (following both from consideration of the action of Lorentz force Eq.\,(7) and from the Larmor theorem) that the net effect of the  magnetic field is the induced circular motion of the paired electrons in the planes transverse to $\textbf{H}$. On the other hand, as it is known for any kind of cyclic motion in magnetic field, the changing $|\textbf{H}|$ changes  the magnitude of $\textbf{v}_i$ only, i. e. the radius of the induced motion  $r$ does not depend of the field provided that $v_i\ll v_0$. 

One can also show that $\widetilde{\textbf{p}}_{cp}$ does not depend on the magnetic field either, namely  
\begin{equation}
	\widetilde{\textbf{p}}_{cp}=\widetilde{\textbf{p}}_1+\widetilde{\textbf{p}}_2=\widetilde{\textbf{p}}_{10}+\widetilde{\textbf{p}}_{20}=0.
\end{equation}

 Hence, the modified London's rigidity holds both in zero and non-zero field. Below we will see that it also holds at non-zero temperature.  

The induced currents in superconductors in the MS (and in other equilibrium S states) are qualitatively identical to the induced bound currents  in regular diamagnetics schematically shown in Fig.\,6. In our case, like in the normal materials, the induced currents mutually compensate  each other in the sample bulk, leaving an uncompensated surface current caused by electrons bound in Cooper pairs with resting centers of inertia. Then the magnetic moment of the sample is exactly the same as the moment produced by a continuous (circumferential) surface current. 

\begin{figure}
	\includegraphics[width=0.7\linewidth]{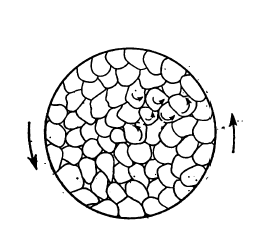}
	\caption{A cross sectional view of a conventional diamagnetic sample showing induced bound currents caused by precession of  the atomic electron orbits; the field is directed into the page (After the textbook of Tamm \cite{Tamm}). This picture is  similar to the induced currents caused by  precession of Cooper pairs in a superconducting sample in the Meissner state. } 
	\label{fig:epsart}
\end{figure}

Now we are ready to calculate magnetic proprieties of our cylindrical sample in the  MS.   

After turning on the applied field $\textbf{H}_0$, the field intensity and the vector potential inside the sample after a short relaxation time become $\textbf{H}$ and $\textbf{A}$, respectively. Then, using Eq.\,(9), we write 
\begin{equation}
	m\textbf{v}_i=-\frac{e}{c}\textbf{A}=-\frac{e}{2c}\textbf{H}\times \textbf{r}.
\end{equation}

Hence,
\begin{equation}
 m\bm{\omega}=-\frac{e}{2c}\textmd{\textbf{H}},
\end{equation}
where $\bm{\omega}=(\textbf{r}\times \textbf{v}_i)/r^2$ is the angular velocity of the induced circular motion of the paired electrons. 

Taking into account that $e<0$, Eq.\,(13) reads that $\bm{\omega}$ is parallel to $\textbf{H}$ and its magnitude is 
\begin{equation}
	\omega=\frac{v_i}{r}=\frac{e}{2mc}H.
\end{equation} 

Comparing with Eq.\,(10) we see that $\omega$ equals the classical Larmor frequency ($g_L=1$).  This confirms that we really deal with the precession of paired electrons with zero total spin, since non-paired electrons (as any other charge) in a magnetic field circulate with a so-called cyclotron frequency $\omega_c=(e/mc)H$.

The induced current per one electron in the pair $J_i$ is
\begin{equation}
	J_i=\frac{e}{2\pi}\omega={\frac{e^2}{4\pi mc}}H.
\end{equation}

As follows from Eqs.\,(14) and (15), neither induced angular velocity $\omega$, nor the induced current $J_i$ depend on $r$. However, this is not the case for the induced magnetic moment and the corresponding change of kinetic energy of the paired electrons: they both depend on $r^2$. On the other hand, $r$ for Cooper pairs with different orientation of $\bm{\mu}_0$ is different. So what we want to know is the mean square $\langle r^2\rangle$,  which we will denote as $r_i^2$. 

Using Eq.\,(15), the magnitude of an average induced magnetic moment per one electron in Cooper pairs is
\begin{multline}
	\mu_i=\frac{1}{c}J_i\pi r_i^2=\frac{e^2r_i^2}{4mc^2}H=\frac{e^2n_s4\pi}{mc^2}\left(\frac{r_i}{2}\right)^2\frac{H}{4\pi n_s}=\\\frac{(r_i/2)^2}{\lambda_L^2}\frac{H}{4\pi n_s}.
\end{multline}
where $n_s$ is a number density of superconducting (i.e. paired) electrons and $\lambda_L$ is so-called London penetration depth defined as
\begin{equation}
	\lambda_L=\left(\frac{mc^2}{4\pi n_s e^2}\right)^{1/2}=\left(\frac{m_{cp}c^2}{4\pi n_{cp} q_{cp}^2}\right)^{1/2},
\end{equation} 
where $m_{cp}=2m$, $q_{cp}=2e$ and $n_{cp}=n_s/2$ are the mass, charge and number density of Cooper pairs, respectively.

Since induced magnetic moments in all Cooper pairs are parallel,  the  magnitude of the magnetic moment of our sample is
\begin{equation}
	M=n_{cp}V\mu_{cp}=\frac{(r_i/2)^2}{\lambda_L^2}\frac{V}{4\pi}H= \frac{(r_i/2)^2}{\lambda_L^2}\frac{V}{4\pi}H_0.
\end{equation}

Comparing this result with Eq.\,(1), we see that the rms radius of the induced motion of the paired electrons is  
\begin{equation}
	r_i=2\lambda_L.
\end{equation}

Since $r$ does not depend on the field, $r_i$ and therefore $\lambda_L$ do not depend on the field either. 
 
Next, basing on Eq.\,(6) one can show that, like in regular diamagnetics, the change of kinetic energy  of  the superconducting electrons $\Delta T$ is the sum of the field-induced kinetic energies of each of these electrons, i.e.
\begin{equation}
	\Delta T=E_k=\epsilon_i n_s V=\epsilon_{cp} n_{cp} V,
\end{equation}
where  $\epsilon_{cp}=2\epsilon_i$ is the average kinetic energy of the induced motion of electrons in one pair. 

Using Eq.\,(14), we write
\begin{equation}
	\epsilon_i=\frac{mv_i^2}{2}=\dfrac{e^2H^2r_i^2}{8c^2m}=\left(\frac{(r_i/2)}{\lambda_L}\right)^2\frac{H^2}{8\pi n_s}.
\end{equation}

Thus, taking into account Eq.\,(19), the  change of kinetic energy of the paired electrons is
\begin{equation}
	\Delta T = \epsilon_i n_sV=\left(\frac{(r_i/2)}{\lambda_L}\right)^2\frac{H^2}{8\pi}V=\frac{H^2}{8\pi}V.
\end{equation}  

In our sample $H=H_0$ and, according to the energy conservation, $\Delta T=E_m=H_0^2V/8\pi$ (see Eq.\,(2)). Hence, our description meets the Law of energy conservation. 

Next, we calculate the gyromagnetic ratio coming from its definition. Using Eqs.\,(14) and (16) we write 
\begin{equation}
	\gamma\equiv\frac{M_i}{L_i}=\frac{\mu_i n_sV}{\iota_i n_s V}=
	\frac{\mu_i}{mv_ir_i}=\frac{e^2r_i^2H}{4mc^2}\frac{2c}{er_i^2H}=\frac{e}{2mc},
\end{equation}
where $M_i$ and $L_i$ are magnitudes of the induced magnetic moment and of the induced angular momentum of the sample, respectively; and $\iota_i$ is an average field-induced angular momentum per one electron.

We see that $\gamma$ is fully consistent with the experimental result of I. Kikoin and Goobar. 

After all, calculate the induction in the sample interior. Using Eqs.\,(16) and (19), and the fact that $\bm{\mu}_i$ is negative, we obtain
\begin{equation}
	\textbf{B}=\textbf{H}+4\pi \textbf{I}=\textbf{H}+4\pi(\bm{\mu}_in_s)=\textbf{H}-4\pi \frac{\textbf{H}}{4\pi n_s}n_s=0,
\end{equation}
in agreement with the result of Meissner and Ochsenfeld and of Rjabinin and Shubnikov.

Correspondingly, the magnetic permittivity $\mu_m(\equiv B/H)$ and susceptibility per unit volume  $\chi$ of the S phase are zero and $-1/4\pi$, respectively.

Naturally, due to microscopic character of the induced currents, there is no problem in establishing the Meissner condition ($B=0$) in samples/domains of any shape, 
as soon as the field $\textbf{H}$  is uniform. The latter is indeed so in ellipsoidal samples, regardless whether they are in the MS or in the inhomogeneous equilibrium states, i.e., in the mixed state of type-II and in the intermediate state of type-I superconductors. 

This explains why the Meissner state  is observed only in the ellipsoidal bodies, as well as the plenty domain shapes observed in the intermediate state.

The above consideration applies to the samples cooled in zero field (ZFC), whereas  the Meissner effect is about both the ZFC and FC (field-cooled) samples. So, what happens in the latter case?

Upon lowering temperature below $T_c(H_0)$ in the fixed field $H_0< H_{c1}(1-\eta)$, a temperature dependent fraction of conduction electrons condenses forming stable Cooper pairs. This means that speed of these electrons drops from $v_F$ to $v_0$, each pair starts orbiting its center of mass and, being in the field, the pairs precess. Like in regular diamagnetics, the latter leads to establishing magnetization $\textbf{I}$, the field intensity $\textbf{H}=\textbf{H}_0-4\pi\eta \textbf{I}$ and the induction $\textbf{B}=\langle \textbf{h} \rangle=\textbf{H}+4\pi\textbf{I}$, where $\langle \textbf{h} \rangle$ is the average microscopic field. Hence, after the short relaxation time needed to establish the intensity $\textbf{H}$, the environment inside the FC sample becomes the same as that in the ZFC sample. Thus, the given description meets the Meissner effect indeed\footnote{The standard description of the Meissner effect is based on the London theory in which this effect is achieved by postulating $B=0$.}.

One more remark. As mentioned above, radius of the electron orbit $R_0$ in precessing Cooper pairs  is fixed (i.e. it does not depend on the field) and the Larmor theorem is exact if $v_i\ll v_0$. Taking typical value of $H_{c1}\sim 100$ Oe and $\lambda_L \sim 10^{-6}$ cm, from Eq.\,(12) one finds $v_i\sim 10^2$ cm/s, which is six orders of magnitude less than $v_0\approx v_F\sim 10^8$ cm/s. So, there is no doubt  that $R_0$ does not depend on the field. On the other hand, since $r_i$ and $R_0$ are quantities of the same or close order of magnitude, this estimate implies that $\omega\ll\omega_0\equiv v_0/R_0$, i.e. the precession is very slow. 

\textbf{Microscopic whirls}. Now, let us reconstruct the current structure of the MS. We understand that all induced currents form  identical circular loops with the rms radius $r_i$  laying in parallel planes transverse to $\textbf{H}$. How these currents are arranged with respect to each other?

Coming from symmetry, one can expect either complete chaos or complete order. Since the system of the ordered currents has lesser free energy, the second option prevails. This is consistent with experimental results of Keesom and Kok \cite{Keesom_Kok_34} that entropy of the sample in the S state is less than that in the N state.

An ordered structure of identical circular currents laying in the planes transverse to $\textbf{H}$ possesses the maximum symmetry if the currents form a 2D hexagonal lattice of cylindrical micro-whirls resembling densely packed and tightly ``wound” micro-solenoids parallel to each other and to $\textbf{H}$.  Length of each whirl/solenoid equals  the sample size along direction of $\textbf{H}$ and its rms diameter is $2r_i=4\lambda_L$. Since the solenoids are parallel, they do not interact. In addition, since the currents are given by the field-induced circular motion of the paired electrons, the complete ordering implies that in all Cooper pairs  this motion is in phase\footnote{Note the direct analogy with the phase of the quantum-mechanical wave function. The wave functions of all Cooper pairs must be in phase since otherwise there would be a total current forbidden by the time reversal symmetry.}. 
 
Just a few facts supporting this statement.  (i) The internal energy of the cylindrical sample in the MS is just the sum of kinetic energies $\epsilon_i$ (Eq.\,(22)) and it does not contain the term(s) responsible for interaction. (ii) In pure type-II superconductors Abrikosov vortices\footnote{The Abrikosov vortices represent holes in the network of the ordered currents of the MS.} make the hexagonal lattice, as was first observed by Essmann and Tr\"{a}uble \cite{Essmann}. (iii) An experimental evidence showing that the Abrikosov vortices do not interact with each other in pure samples \cite{MS}.  

Naturally, the complete ordering of the field-induced currents together with Eq.\,(6) mean that entropy of the ensemble of Cooper pairs $S_{cp}$ is zero, exactly as it takes place for the magnetic part of entropy in regular diamagnetics. Therefore the given theoretical interpretation, referred to as a micro-whirls (MW) model, is consistent with the absence of the thermoelectric effects in superconductors and justifies the postulate of zero entropy of superconducting electrons of the two-fluid model of Gorter and Casimir. 

One more detail about the micro-whirls. A spacing between current loops $\Delta$ along the whirl axis is a universal quantity\footnote{Since $r_i^2=(2\lambda_L)^2\sim 1/n_{cp}$, the universality of $\Delta$ is a necessary condition that the whirls, i.e., ordered Cooper pairs, fill the entire volume of the sample in the MS at any change of $n_{cp}$ (occurring when $T$ changes) regardless of the specific material of the sample.} equal to 
\begin{equation}
	\Delta=\frac{2e^2}{mc^2}=5.6\cdot 10^{-13} cm \approx 6\,fm.
\end{equation}    
 
 So $\Delta$ is about a tripled size of a proton (1.7 $fm$) and therefore the whirls are similar to very tightly wound solenoids indeed. 

An important feature of a single whirl is that it is characterized by two microscopic quantities with dimension of length. Those are the radius of the orbital motion $R_0$ and the radius of the field-induced circular motion $r_i$. As mentioned, $r_i$ is proportional to $R_0$, both radii do not depend on the field and the ratio $r_i/R_0$ is a constant of superconducting material. However, if vector $\textbf{r}_i$ always lays in the transverse plane, $\textbf{R}_0$ is symmetrically distributed over 3D space as follows from Eq.\,(6). So, like in the Langevin theory of diamagnetism, a mean squared projection of $\textbf{R}_0$ on the transverse plane $R_\perp^2$ equals $2R_0^2/3$. In other words, $R_\perp$ is an rms radius of a cylinder coaxial to the cylinder with radius $r_i$ and filled by orbiting Cooper pairs. Hence, an individual whirl can be viewed as a ``double wall" cylinder with radii $R_\perp$ and $r_i$. 

A parameter of material in the MW model is $\aleph$ (aleph) defined as 
\begin{equation}
	\aleph=\frac{r_i}{R_\perp}.
\end{equation}

\begin{figure}
	\includegraphics[width=0.8\linewidth]{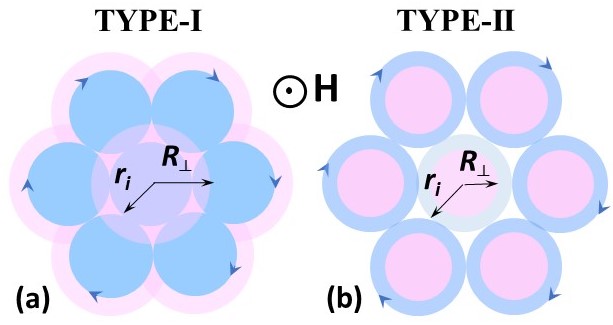}
	\caption{Schematics of the micro-whirl structure in the transverse cross section of type-I (a) and type-II (b) superconductors. In (a) $r_i<R_\perp$ or $\aleph<1$; in (b) $r_i>R_\perp$ or $\aleph>1$. The induced currents (all are in phase with each other) are designated by arrows; these induced currents are in phase in all Cooper pairs.  Areas filled with Cooper pairs are colored in pink; the cross-sectional areas of the induced currents are colored in blue. $R_\perp$ is the rms projection of the pairs radius $R_0$ onto the plane perpendicular to $\textbf{H}$; $r_i$ is the rms radius of the induced motion of the paired electrons. Boundaries of all cylinders are diffuse (unsharp), the cylinders overlap without leaving voids. The field $\textbf{H}$ is directed toward the reader.(After V. Kozhevnikov, JSNM \textbf{34}, 1979 (2021).)} 
	\label{fig:epsart}
\end{figure}

In materials with $\aleph<1$ the S/N surface tension is positive and correspondingly such materials represent type-I superconductors. In type-II superconductors the S/N interphase energy is negative and $\aleph>1$. The cross section of the whirl structure in type-I and type-II superconductors is shown in Fig.\,7. 

 \textbf{Penetration depth.} Another important property of the MS is the penetration depth $\lambda$, a width of the surface layer  where $B$ decades from $B_{ext}(=B_0=H_0$ for the cylindrical samples) to zero in the sample interior. 

According to the London theory, in a sample with a plane boundary (i.e., in the massive cylindrical samples),  the induction $B(z)$, where $z$ is the distance from the surface, decays  exponentially with the decay constant $\lambda_L$. Correspondingly, $\lambda$ is infinite and an effective penetration depth $\lambda_{eff}=\lambda_L$\footnote{$\lambda_{eff}$ is such a depth over which the flux of the penetrating field is the same as the flux of the real field and the flux density $B$ is constant and equal to $B_{ext}$.}. What is $\lambda$ in the MW model?

To answer this question one needs to know the angular distribution of $\bm{\mu}_0$, which is beyond the scope of the semi-classical consideration employed in the  model. One can estimate $\lambda$ assuming that the boundary of the blue cylinders in Fig.\,7 is sharp or the induced current $J_i$ is linear. Then for the plane sample boundary $\lambda=r_i=2\lambda_L$ and $\lambda_{eff}=\lambda_L/2$. This is, of course, an oversimplified estimate, however in any case in the MW model $\lambda$ is finite and close to $\lambda_L$.        

\textbf{Temperature.} All properties discussed so far are related to those for the samples at $T=0$. However, it is quite obvious that nothing will change if $T\neq0$.

Indeed, we have seen that entropy  of the ensemble of Cooper pairs $S_{cp}$ is zero. Therefore, according to the Third law (Nernst's theorem), the temperature of the ensemble of Cooper pairs $T_{cp}$ is zero as well, regardless on the sample temperature $T$.  Hence, all results obtained in this section hold in  the whole temperature range of the existence of Cooper pairs. In other words, the paired electrons are in the ground state in the entire temperature and field range of the S state.

However, $T_{cp}=0$ does not mean that the sample temperature has no effect on the properties of paired electrons, since otherwise $T_c$ would be infinite. The sample temperature affects the lattice polarization responsible for the electron pairing. Correspondingly, it changes $n_s(=2n_{cp})$, as shown in the two-fluid model. Therefore, the change of $T$ leads to the change of $\lambda_L$ and, correspondingly, to  the change of $r_i$, the rms radius of the field-induced motion of electrons in the pairs. On the other hand, $r_i$ is proportional to $R_0$ with the proportionality coefficient $\aleph$ (more strictly $\sqrt{2/3} \aleph$) determined by the superconducting material. In its turn, $\aleph$ does not depend on $T$ due to the constancy of $T_{cp}(=0)$.

\section*{FLUX QUANTIZATION} 

\begin{figure}
	\includegraphics[width=0.8\linewidth]{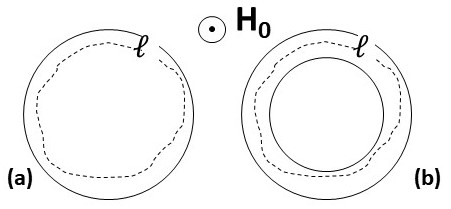}
	\caption{Cross section of solid (a) and hollow (b) cylindrical samples in a field $\textbf{H}_0$  parallel to the long axis of the samples. $l$ is a loop lying away from the wall(s) of the cylinders.  } 
	\label{fig:epsart}
\end{figure}

Let us consider a closed macroscopic loop $l$ inside a sample in the MS away from its surface. For simplicity,  take a cylindrical sample and the loop lying in the transverse plane as shown in Fig.\,8a. Now, calculate the circulation of the linear momentum $\widetilde{\textbf{p}}_{cp}$ over this loop.  

Moving along the loop, we will pass through lots of pairs, so we should consider the average momentum $\langle\widetilde{p}_{cp}\rangle$  and the average quantum number $\langle n\rangle$. Since all Cooper pairs are in identical conditions $\langle n\rangle =n$. Therefore, the Bohr-Sommerfeld quantization condition for this case is

\begin{equation}
\oint_l \langle\widetilde{\textbf{p}}_{cp}\rangle\cdot d\textbf{l}=nh,
\end{equation}
where $l$ is the loop length.

Now, open $\langle\widetilde{\textbf{p}}_{cp}\rangle$ and take into account that in the MS $n=0$. Then, Eq.\,(27) becomes
\begin{multline}
\oint_l \langle\widetilde{\textbf{p}}_{cp}\rangle\cdot d\textbf{l}=\\\oint_l \langle m\textbf{v}_{i1}\rangle\cdot d\textbf{l}+\oint_l \langle m\textbf{v}_{i2}\rangle\cdot d\textbf{l}+\frac{2e}{c}\oint_l \langle \textbf{A}\rangle \cdot d\textbf{l}=0.
\end{multline}

The first two integrals are zero due to mutual compensation of the induced kinetic linear momentums of electrons in neighboring pairs, similar as it takes place in the regular diamagnetics. 

Now, what is  $\langle \textbf{A}\rangle$, the average of the vector potential of the intensity $\textbf{H}$? To answer, we apply Stokes' theorem; then Eq.\,(28) is rewritten as
\begin{equation}
\frac{2e}{c}\oint_l \langle \textbf{A}\rangle \cdot d\textbf{l}=\frac{2e}{c}\int_{F_l}(\nabla \times\langle \textbf{A}\rangle)\cdot d\textbf{f}=0,
\end{equation}
where $F_l$ is the area of  a surface bounded by the loop $l$ and $d\textbf{f}$ is a vector element of this surface. 

From Eq.\,(29) we see that the integral over the area $F_l$ is the flux of a vector $[\nabla\times\langle \textbf{A}\rangle]$  and this flux equals zero. Therefore, since $\nabla\times \textbf{A}\equiv\textbf{H}\neq0$, $\langle \textbf{\textbf{A}}\rangle\neq \textbf{A}$.

On the other hand, inside our sample the induction $\textbf{B}$ and therefore its flux is zero (Eq.\,(39)). Therefore,  Eq.\,(29) suggests that $\langle \textbf{A}\rangle$ is the vector potential of the magnetic flux density (induction) $\textbf{A}_B$ defined as $\textbf{B}=\nabla\times \textbf{A}_B$. In other words, the vector potential of the flux density $\textbf{A}_B$ is a macroscopic average of the vector potential $\textbf{A}$ determining the field induced microscopic currents $J_i$. This is an exact match with the classical definition of $\textbf{A}_B$  as a macroscopic mean of the vector potentials caused by the microscopic currents.   So, putting $\langle \textbf{A}\rangle=\textbf{A}_B$, we rewrite  Eq.\,(29) as  
\begin{equation}
\frac{2e}{c}\int_{F_l}(\nabla \times\langle \textbf{A}\rangle)\cdot d\textbf{f}=\frac{2e}{c}\int_{F_l} \textbf{B}\cdot d\textbf{f}=\frac{2e}{c}\varPhi=0,
\end{equation}
where $\varPhi$ is the magnetic flux through the area $F_l$.

Now, take a hollow (tube-like) thick-wall long cylinder shown in Fig.\,8b, apply the field $H_0(<H_{c1})$ parallel to its longitudinal axis and cool the cylinder below $T_c$. The loop \textit{l} encircles the opening of the cylinder and lies inside the wall far (compare to $\lambda$) from the inner and outer cylinder surfaces. The induction inside the wall is zero, implying that, as in Eq.\,(28), $\langle m\textbf{v}_i\rangle=0$. On the other hand, the flux $\varPhi$ inside the hollow cylinder is frozen and therefore it is not zero. Then Eq.\,(27) yields

\begin{multline}
\oint_l \langle\widetilde{\textbf{p}}_{cp}\rangle\cdot d\textbf{l}=\frac{2e}{c}\oint_l \langle \textbf{A}\rangle \cdot d\textbf{l}=\frac{2e}{c}\int_{F_l}\nabla \times \langle \textbf{A}\rangle\cdot d\textbf{f}=\\\frac{2e}{c}\int_{F_l} \textbf{B}\cdot d\textbf{f}=\frac{2e}{c}\varPhi=nh.
\end{multline} 

Hence, the magnetic flux passing through the opening (plus surrounding it penetration area) in a multiply connected superconductor is 
\begin{equation}
\varPhi=\frac{c}{2e}nh.
\end{equation}

This is the famous London's flux quantization but for the paired electrons. Hence, the origin of the superconducting flux quantization is the Bohr-Sommerfeld quantization condition, which also justifies existence of the field induced persistent currents in  the MS.  

Eq.\,(32) implies that the flux quantum and therefore the flux passing through each so-called Abrikosov vortex in type-II superconductors in the mixed state is
\begin{equation}
\varPhi_0=\frac{c}{2e}h=\frac{\pi c \hbar}{e}.
\end{equation}  

As well known, the flux quantization Eq.\,(32) and the single flux quantum in the Abrikosov vortices Eq.\,(33) are in full agreement with experiment. 

Thus, inside the S phase the vector  potential $\textbf{A}_B=\langle\textbf{A}\rangle$, where $\textbf{A}$ is the vector potential of the field $\textbf{H}$.  Now, what is the value of $\textbf{A}_B$ inside the sample in the MS? 

In the plane perpendicular to $\textbf{H}$ we have uniformly distributed identical induced  circular currents, i.e. in-plane currents with the same magnitude $J_{cp}$ and the same rms radius $r_i$ circulating clockwise relative to $\textbf{H}$. Therefore, exactly as it takes place in regular diamagnetics, equal amount of electricity flows in opposite directions throughout an out-of-plane cross section of an arbitrary chosen volume element $dV$. Hence, an average current density $\langle\textbf{j}\rangle=en_s\langle\textbf{v}_i\rangle$ is zero in any volume element of the sample interior or inside the S phase. So, taking into account Eq.\,(9), we write 
\begin{equation}
\textbf{A}_B\equiv\langle \textbf{A}\rangle=\langle \textbf{v}_i\rangle=\langle \textbf{j}\rangle=0.
\end{equation}
    
Thus, inside the sample in the MS, or in general inside the S phase, in spite of non-zero $\textbf{H}$, an average of its vector potential $\langle \textbf{A}\rangle$, as well as the average induced velocity $\langle \textbf{v}_i\rangle$, and the average induced current $\langle \textbf{j}\rangle$ are all equal zero. Out of these three equalities the last one [$\langle \textbf{j}\rangle=0$] is the most important: it shows that all currents in the S phase are mutually compensated, as required by the Law of the time reversal symmetry.

\section*{TOTAL CURRENT}

As we have seen, the equilibrium magnetic properties of superconductors are qualitatively similar to the properties of conventional diamagnetics. In both cases these properties are due to magnetization arising from precession of the microscopic magnetic moments caused by the orbital motion of electrons bound either in atoms (conventional diamagnetics) or in Cooper pairs (superconductors). A colossal quantitative  difference in the magnetic susceptibilities in these materials (up to 5 orders of magnitude!) is due to the differences in the size of the orbits and correspondingly in the radii of the induced microscopic currents of the bound electrons. 

However, there is also an important qualitative difference in these induced currents. Namely, in conventional diamagnetics the orbiting electrons are bound in motionless atoms (fixed in the crystal lattice), while in superconductors - in movable Cooper pairs. The pairs mobility allows them to compose the whirl structure with zero entropy.  On the other hand, since the pairs have electric charge, they can form an electric current, referred to as the total current. It includes the transport current and the field-induced current encircling the openings in multiply connected bodies. 
In the latter case, as was for the first time observed by Deaver and Fairbank  \cite{Deaver}, the total current is quantized due to the flux quantization.

The total current in superconductors plays a role similar to that of the transport current  in conventional metals, where it is executed by conduction electrons. The latter obey Fermi-Dirac statistics and their energy equals the Fermi energy $E_F$. This implies that the carriers of the transport current in normal metals, or in the N phase of superconductors, are  "very hot": their temperature, the temperature of the ensemble of conduction electrons, equals the Fermi temperature $T_F=E_F/k_B\sim 10^4$ K ($k_B$ is the Boltzmann constant), i.e., it is about the same as the sun surface temperature.

In striking contrast, the charge carriers (Cooper pairs) of the total current in the S phase are "deadly cold". Since their  spins are zero, the pairs obey Bose-Einstein statistics and, since the temperature of the ensemble of  the paired electrons $T_{cp}$ is zero, they form the Bose-Einstein condensate (BEC).  This (zero temperature) leads to the disappearance of the thermoelectric effects, as observed in experiments. The amazing transformation of the ``very hot" single electrons to the ``zero-temperature" electron pairs can be compared with the fact known from the relativity theory: two flying apart massless photons form a  massive pair located in their center of mass.

Therefore, the total current in superconductors represents the transport current in BEC, known on the properties of superfluid helium. Correspondingly, the total current set in a closed superconducting circuit continues running without energy dissipation provided speed of the charge carriers (density of the total current) is lesser of a definite critical value \cite{Landau_41,Feynmann_55}.

Thus, in the MW model the electromagnetic properties of superconductors caused by the total current\footnote{We are talking about the dc total current. Consideration of the ac current must include contribution of the unpaired electrons, which is significant at frequencies $\gtrsim10^3$ MHz. }  
resemble  the properties of hypothetical perfect conductors. 
 
The hell-mark of the perfect conductors is irreversibly of the sample magnetic moment induced by the changing applied magnetic field; its direction is determined by the Lenz law whereas the magnetic moment caused by magnetization is always diamagnetic. On this reason the multiply connected bodies can never be in thermodynamic equilibrium and therefore neither the principal of the free energy minimum nor the time reversal symmetry can be applied to them.  

However, it should be stressed that even in the presence of the total current, magnetic properties of superconductors in the MW model are different from those of the perfect conductors. Since the total current is accompanied by its own magnetic field, it represents a combination of the whirl and  translational motion of the paired electrons regardless of the presence or absence of the applied field. This explains the observation reported by Meissner and Ochsenfeld in their fundamental paper of 1933 \cite{Meissner}:  
``When the parallel superconductors are connected end-to-end in series and an external current is connected to flow through them above the critical temperature the magnetic field between the superconductors is increased below the transition temperature the external current being unchanged".

\section*{SUMMARY AND CONCLUSIONS}

In this chapter we presented the basics of the micro-whirls model of superconductivity and a description of main electrodynamic properties as it follows from the model. It is important to underline, that the ``big zeroes" of superconductivity listed in the abstract, as well as the zero generalized linear momentum (modified London's rigidity), the zero average vector potential and the zero average current in the sample bulk, all rise from a single root: quantization of the angular momentum of electrons combined into Cooper pairs. It will not be redundant to note that the fact of zero temperature of Cooper pairs regardless of the sample temperature means that there is no fundamental limit in the critical temperature of superconductors, as it is really observed after the discovery of high-temperature superconductors by Bednorz and M\"{u}ller in 1986 \cite{Muller}.
\vspace{5 mm}

\textbf{Acknowledgments}. The author is deeply grateful to Prof. Vladimir Kresin and Prof. Orest Simko for valuable comments after reading the manuscript. 

\section*{Further Reading}

Details of the micro-whirls model with necessary references to experimental and theoretical works, and a broader description of the superconducting properties are available in 

V. Kozhevnikov, ``Meissner Effect: History of Development and Novel Aspects",  J. Supercond. Nov. Magn. \textbf{34}, 1979–2009 (2021).

A perfect review of experimental and theoretical works up to 1952 is available in 

D. Shoenberg, \textit{Superconductivity}, 2nd. ed., (Cambridge, University Press, 1962).

For fundamentals of electrodynamics we recommend

I. E. Tamm, \textit{Fundamentals of the Theory of Electricity} 9th ed. (Mir, Moscow, 1979),

E. M. Purcell, \textit{Electricity and Magnetism}, 2nd ed. (McGraw-Hill, Boston, 1985),

L. D. Landau, E.M. Lifshitz and L. P. Pitaevskii, \textit{Electrodynamics of Continuous Media}, 2nd ed. (Elsevier, Oxford, 1984).

These textbooks complement but not replace one another. 

Thermodynamics of superconductors is discussed in 

V. Kozhevnikov, \textit{Thermodynamics of Magnetizing Materials and Superconductors} (CRC Press, Boca Raton, 2019). 

\section*{References}


\end{document}